\documentclass[aps,prb,twocolumn,showpacs]{revtex4}
\usepackage[utf8]{inputenc}
\usepackage{graphicx}
\usepackage[pdftex,hypertexnames=false,pdfpagemode=UseThumbs,linkbordercolor={1 1 1}, citebordercolor={1 1 1},pdfstartview=FitH]{hyperref}
\usepackage{amsmath}
\usepackage{wasysym}
\usepackage{diagbox}
\usepackage{bbold}
\usepackage{tikz}
\DeclareMathAlphabet{\mathpzc}{OT1}{pzc}{m}{it}
\newcommand{\underdash}[1]{\tikz[baseline=(todotted.base)]{\node[inner sep=1pt,outer sep=0pt] (todotted) {#1};\draw[densely dashed] (todotted.south west) -- (todotted.south east);}}
\renewcommand{\vec}[1]{\boldsymbol{#1}}
\begin{document}
\title{Tuning antiferromagnetism of vacancies with magnetic fields in graphene nanoflakes}
\author{Matthias Droth}
\author{Guido Burkard}
\affiliation{Department of Physics, University of Konstanz, 78457 Konstanz, Germany}
\begin{abstract}
Graphene nanoflakes are interesting because electrons are naturally confined in these quasi zero-dimensional structures, whereas confinement in bulk graphene would require a bandgap. Vacancies inside the graphene lattice lead to localized states and the spins of such localized states may be used for spintronics. We perform a tight-binding description of a nanoflake with two vacancies and include a perpendicular magnetic field via a Peierls phase. The tunnel coupling strength and from it the exchange coupling between the localized states can be obtained from the energy splitting between numerically calculated bonding and antibonding energy levels. This allows us to estimate the exchange coupling $J$, which governs the dynamics of coupled spins. 
We predict the possibility of switching in-situ from $J>0$ to $J=0$ by tuning the magnetic field. In the former case, the ground state will be antiferromagnetic with Néel temperatures accessible by experiment.
\end{abstract}
\maketitle
\section{Introduction}
Beyond the outstanding mechanical, optical, and electronic characteristics common to bulk graphene \cite{Novoselov2004,Lee2008,Nair2008,Kuzmenko2008,Eda2008,Lin2010}, graphene nanoflakes are predicted to feature magnetic properties, as well \cite{FernandezRossier2007,Ezawa2007,Yazyev2010}. These qualities make such graphene nano-islands very interesting for spintronics and other applications \cite{Petta2005,Trauzettel2007,Hong2013}. Lattice defects can occur due to chemisorption of hydrogen molecules, but they can also be generated on purpose by means of ion or electron beam irradiation \cite{Yazyev2007,Krasheninnikov2007,Robertson2013}. Such defects are expected to give rise to magnetic moments of about 1 Bohr magneton. The associated magnetic ordering can in principle be ferromagnetic as well as antiferromagnetic \cite{Lieb1989,Yazyev2007,Palacios2008,Uchoa2008,Grujic2013} and recent progress in spin sensitive measurements allows one to probe these predictions \cite{Wiesendanger2009,Decker2013,Nair2012}. In practice, however, modifying the magnetic properties of defect-induced magnetic graphene typically requires the preparation of new devices. Graphene nanoflakes can be grown using chemical vapor deposition (CVD), typically with zigzag boundaries and hexagonal symmetry of the entire flake \cite{Luo2011,Phark2011,Subramaniam2012}. Hexagonal nanoflakes with armchair boundaries can be constructed bottom-up from aromatic molecules \cite{Wu2007, Su2009}. In addition, it has been reported that the interaction of the nano-island edges with the substrate smoothes the boundary and enhances the symmetry of the electronic wave functions \cite{Hamalainen2011}. In analogy to the hydrogen molecule, two localized states in a double quantum dot (DQD) can hybridize to form bonding and antibonding eigenstates of the combined system. In return, the localized states can be obtained by taking the even and odd superpositions of bonding and antibonding states. The exchange coupling $J$ describes the coupling between the two localized spins \cite{Burkard2006,Schrieffer1966}. In this article, we calculate $J$ as a function of the magnetic field and for different flake configurations.
\begin{figure}[t!]\centering\includegraphics[width=0.455\textwidth]{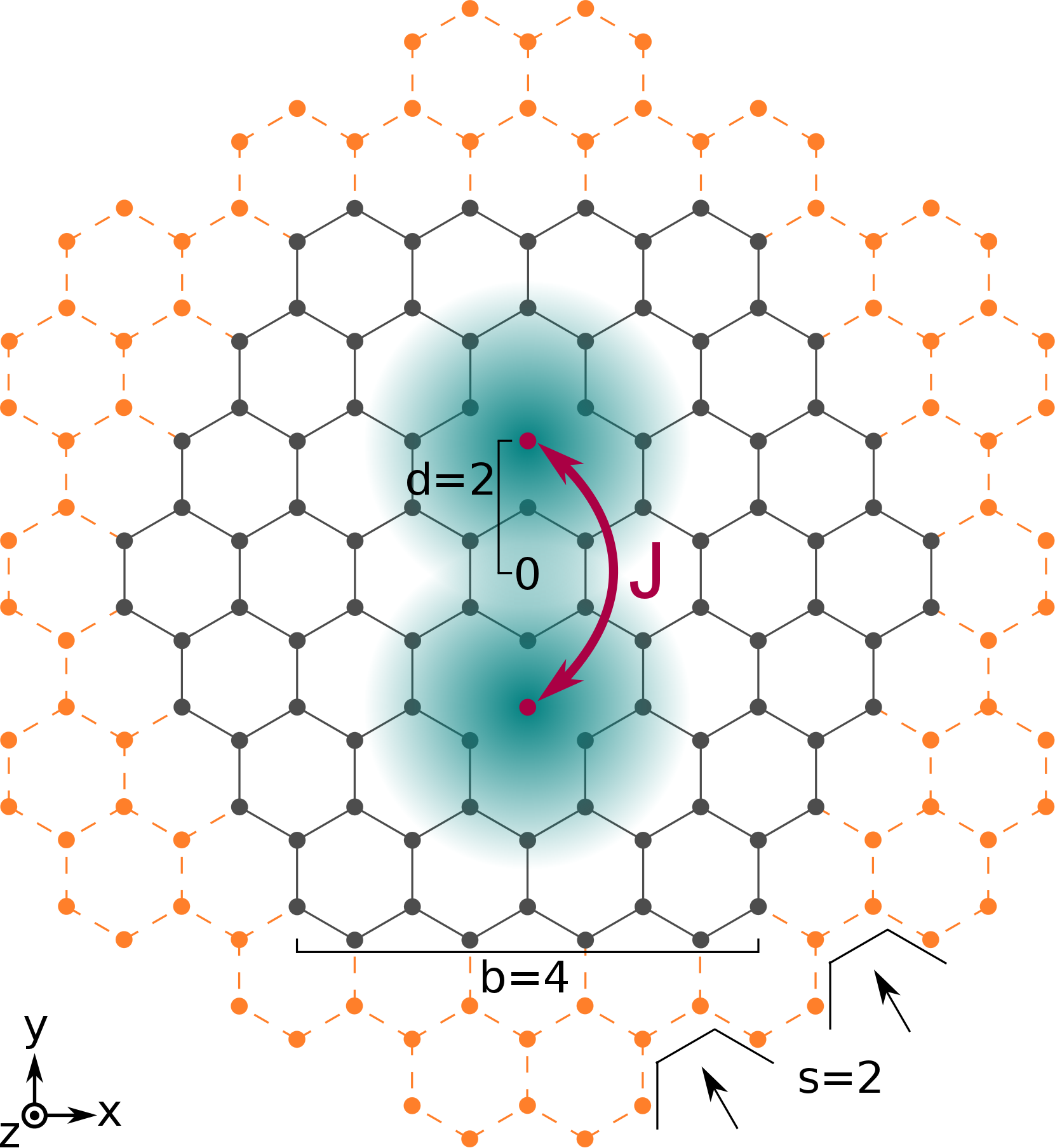}\caption{(Color online) Schematic of a hexagonal graphene nanoflake with zigzag (armchair) boundaries. Grey (and orange) dots --- connected by solid (and dashed) lines --- indicate the locations of carbon atoms. The zigzag (armchair) terminated flake is specified by the number of benzene rings (armchair sections) along each edge, $b$ ($s$), and the distance $d$ (in units of the atomic distance $a=1.42{\rm \AA}$) of the vacancies, located at $\vec{r}_{\rm vac}=(0,\pm y)$, from the Cartesian origin in the flake center. The sketched island has a $(b=4,d=2)$ $[(s=2,d=2)]$ configuration. The vacancies (red dots) give rise to localized spin states (green shade) whose mutual dynamics is described by the exchange coupling $J$. A magnetic field $\vec{B}\parallel\vec{e}_z$ can be applied perpendicularly to the flake plane.}\label{pic1}\end{figure}

A typical graphene nanoflake with zigzag (or armchair) edges, hexagonal symmetry, and two lattice vacancies is sketched in Fig.~\ref{pic1}. Each vacancy gives rise to localized states and thus serves as a quantum dot \cite{Pereira2006,Xiong2007}. The entire flake with two vacancies is therefore a realization of a DQD. If the vacancies are located at positions $\vec{r}_{\rm vac}=(0,\pm y)$, the flake retains some symmetry which, in our case, also applies to the probability densities of the electronic states. The complete eigensystem is found by numerical diagonalization of a tight-binding Hamiltonian where nearest neighbors up to third order can be taken into account and a perpendicular magnetic field $\vec{B}\parallel\vec{e}_z$ is included via a Peierls phase. Interactions are effectively taken into account in a second step when calculating $J$. The retained symmetry allows us to superpose the eigenstates in a meaningful way. We calculate the exchange coupling $J$ as a function of the magnetic field and for different zigzag (armchair) flake configurations, which we specify by the number of benzene rings (armchair sections) per edge, $b$ ($s$), and the distance between the vacancies and the flake center, $d$, as shown in Fig.~\ref{pic1}. We find that $J$ can be tuned over several orders of magnitude within one device and can even vanish for certain flake configurations by changing the magnetic field. For finite $J$, the ground state of the system is antiferromagnetic. The according Néel temperature depends on the flake geometry and ranges from below $4\,$K to values beyond room temperature. That is, our results are in reach of experimental analysis via spin-polarized scanning tunneling microscopy or SQUID magnetometry \cite{Wiesendanger2009,Decker2013,Nair2012}.
\section{Tight-Binding Model}\label{tbmSec}
We consider a tight-binding Hamiltonian with hopping between neighbors up to third order,
\begin{eqnarray}
\mathcal{H}=\sum_{\langle i,j\rangle}t^{(1)}_{ij}c^{\dagger}_ic_j+\sum_{\langle\langle i,j\rangle\rangle}t_{ij}^{(2)}c^{\dagger}_ic_j+\sum_{\langle\langle\langle i,j\rangle\rangle\rangle}t^{(3)}_{ij}c^{\dagger}_ic_j\,,\label{hamil}
\end{eqnarray}
where the hopping from atom $j$ to a neighbor of $n$-th order, $i$, depends on the magnetic field $\vec{B}\parallel\vec{e}_z$ via the Peierls phase,
\begin{eqnarray}
t_{ij}^{(n)}(B)&=&t_{ij}^{(n)}(0)\,{\rm exp}\!\left[i\frac{e}{\hbar}\int_{\vec{R}_i}^{\vec{R}_j}\vec{A}(\vec{r})\cdot\text{d}\vec{r}\right]\nonumber\\
&=&t_{ij}^{(n)}(0)\,{\rm exp}\!\left[i\frac{eB}{2\hbar}(y_i+y_j)(x_i-x_j)\right]\,.\label{phase}
\end{eqnarray}
We use the Landau gauge $\vec{A}(\vec{r})=-By\vec{e}_x$ and zero field hopping amplitudes $t_{ij}^{(1)}(0)=2.8\,\text{eV}$, $t_{ij}^{(2)}(0)=0.7\,\text{eV}$, and $t_{ij}^{(3)}(0)=0.3\,\text{eV}$. The operator $c^{\dagger}_i$ ($c_i$) creates (annihilates) an electron at site $\vec{R}_i$. At zero magnetic field, the symmetries of the Hamiltonian are the same as the lattice symmetries, as seen in Fig.~\ref{pic1}: the mirror symmetries $\mathcal{M}_x\!:x\mapsto-x$ and $\mathcal{M}_y\!:y\mapsto-y$ as well as the rotation by $\pi$, $\mathcal{R}_2\!:(x,y)\mapsto(-x,-y)$. At finite fields only the twofold rotation $\mathcal{R}_2$ remains.

The numerically obtained eigenstates have an arbitrary phase. However, we find that it is possible to multiply any eigenstate $|n\rangle$ with a phase such that $\langle\vec{r}|n\rangle=\langle n|\mathcal{M}_x\vec{r}\rangle$. While the probability density $|\langle\vec{r}|n\rangle|^2$ remains unaffected, these phase rotations do matter for the probability densities of even and odd superpositions of two eigenstates. In order to obtain states localized at $\vec{r}_{\rm vac}$ by forming these superpositions, it is necessary to perform these phase rotations on the \mbox{(anti-)}bonding eigenstates.
\section{Localized States and Exchange Coupling}
The graphene nanoflake with vacancies can be interpreted as a symmetric, unbiased DQD. Such a system can be described by the Hamiltonian
\begin{eqnarray}
\mathcal{H}_{\rm DQD}=\begin{pmatrix}\bar{E}&t\\t^*&\bar{E}\end{pmatrix}\,,\label{DQD}
\end{eqnarray}
where the localized states $\{|\!+\!y\rangle,|\!-\!y\rangle\}$ form the basis, $t$ is the hopping amplitude from site to site, and $\bar{E}$ is the degenerate eigenenergy for $t=0$. An arbitrary gauge is taken into account via the phase $\phi$, that is, $t=|t|e^{i\phi}$. The eigensystem of Eq.~(\ref{DQD}) is
\begin{eqnarray}
E_{\pm}&=&\bar{E}\pm|t|\,,\label{barEt}\\
|\psi_{\pm}\rangle&=&(|\!+\!y\rangle\pm e^{-i\phi}|\!-\!y\rangle)/\sqrt{2}\,.\label{estate}
\end{eqnarray}
Thus, the hybridized {\it bonding} ($|\psi_-\rangle$) and {\it antibonding} ($|\psi_+\rangle$) states are superpositions of the localized states and their energy splitting is given by $\Delta=2|t|$.

The diagonalization of the tight-binding Hamiltonian Eq.~(\ref{hamil}) yields the bonding and antibonding eigenstates and the according energy spectrum. If two states $|\psi_{\pm}\rangle$, bonding and antibonding, are selected, then the according localized states $|\!\pm\!y\rangle$ are obtained by superposing these \mbox{(anti-)}bonding states. This corresponds to undoing the superposition in Eq.~(\ref{estate}). The magnitude of the hopping between the localized states is easily obtained from the energy splitting between the \mbox{(anti-)}bonding states: $|t|=\Delta/2$. To do this, we need to select a pair of bonding and antibonding states and superpose them after rotating them with phases as described at the end of Sec.~\ref{tbmSec}.

We find that if hopping amplitudes beyond nearest neighbors are taken into account and $B$ is finite \cite{Bfield}, no degeneracies  occur (except for spin, which will only be considered later). Since $\mathcal{R}_2$ commutes with the Hamiltonian Eq.~(\ref{hamil}), the energy eigenstates are also eigenstates of $\mathcal{R}_2$, with eigenvalues $+1$ (even) and $-1$ (odd). We consider any two states $|n\rangle$, $|m\rangle$ with (i) eigenenergies that lie next to each other in the discrete energy spectrum, $E_n=E_{m-1}$, and with (ii) opposite symmetry under the twofold rotation, $\langle n|\mathcal{R}_2|n\rangle\langle m|\mathcal{R}_2|m\rangle=-1$. We refer to the lower energetic state as {\it bonding} and the higher energetic one as {\it antibonding}, see Eq.~(\ref{estate}). In addition to $\mathcal{R}_2$, the lattice also possesses the symmetries $\mathcal{M}_x$ and $\mathcal{M}_y$. We find that the probability density of any eigenstate, $|\langle\vec{r}|n\rangle|^2$, also possesses these symmetries. Since the localized states are localized in the upper/lower half of the flake, their probability densities should only possess the symmetry $\mathcal{M}_x$. This symmetry fixes the relative phase in the superposition of the bonding and antibonding states.

The procedure described so far allows us to find the localized states $|\!\pm\!y\rangle$ for any selection of \mbox{(anti-)}bonding states. To describe the spin physics in the DQD, we include spin $\sigma=\uparrow,\downarrow$ and an on-site Coulomb repulsion $U$. It is well known that the system has six possible states: three spin triplets and three spin singlets \cite{Burkard2006}. In the weak tunneling regime $|t|\ll U$, the triplet state $|T_0\rangle=\frac{1}{\sqrt{2}}(c^{\dagger}_{+y\uparrow}c^{\dagger}_{-y\downarrow}+c^{\dagger}_{+y\downarrow}c^{\dagger}_{-y\uparrow})|0\rangle$ and singlet state $|S\rangle=\frac{1}{\sqrt{2}}(c^{\dagger}_{+y\uparrow}c^{\dagger}_{-y\downarrow}-c^{\dagger}_{+y\downarrow}c^{\dagger}_{-y\uparrow})|0\rangle$ decouple from the other states and are effectively described by the Hamiltonian \cite{Burkard2006,Schrieffer1966}
\begin{eqnarray}
\mathcal{H}_{\rm TS}\approx
\begin{pmatrix}0&0\\0&-J\end{pmatrix}\,,\quad J=\frac{4|t|^2}{U}\,,\label{Heff}
\end{eqnarray}
where the basis is $\{|T_0\rangle,|S\rangle\}$ and the Coulomb repulsion is $U=e^2/4\pi\epsilon_0|r|$, with the elementary charge $e$ and the vacuum permittivity $\epsilon_0$. For $|r|$, we use the standard deviation of the probability density of the corresponding localized state. The Zeeman term $g\mu_B\vec{B}\cdot\vec{\Sigma}$, where $g$ is the electron $g$ factor in graphene, $\mu_B$ is the Bohr magneton, and $\vec{\Sigma}=\vec{\sigma}_{+y}+\vec{\sigma}_{-y}$ is the total spin, commutes with $\mathcal{H}$ as well as $\mathcal{H}_{\rm TS}$ and hence does not affect the calculation of $J$.
\section{Results}
Since the nanoflake consists of a total number of $N$ atoms, the tight-binding Hamiltonian in Eq.~(\ref{hamil}) has dimension $N\times N$. Because of spin degeneracy, the $N$ sorted eigenenergies $E_n$ are only filled up to $E_{N/2}$ (counting from the bottom of the spectrum) by the $p_z$ electrons. To simplify our notation, we now count eigenstates and eigenenergies with respect to the middle of the spectrum. That is, instead of $E_{N/2+n}$ we will just write $E_n$ and we set $E_0{=}0$.

We calculate the exchange coupling $J$ for three fundamentally different situations: (i) $t^{(1)}_{ij}{\neq}t^{(2)}_{ij}{=}t^{(3)}_{ij}{=}0$ in Eq.~(\ref{hamil}) with armchair terminated flakes and (ii) zigzag terminated flakes, as well as (iii) $t^{(n)}_{ij}{\neq}0$ ($n{=}1,2,3$) with zigzag boundaries. For (i) and (ii), the localized vacancy states lie in the middle of the symmetric energy spectrum \cite{Pereira2006,Nair2012}. Zigzag edges are energetically favored and hence more likely to occur in nanoflakes grown by CVD. For (iii), the localized states do not necessarily lie in the middle of the energy spectrum, which makes their identification non-trivial.
\subsection{Armchair and zigzag terminated flakes with hopping up to first nearest neighbors}
For $t^{(n)}_{ij}{=}0$ ($n{>}1$), the on-site wave functions of eigenstates $|0{-}n\rangle$ and $|1{+}n\rangle$ (i.e.~states that lie symmetrically with respect to the middle of the energy spectrum) differ only by a phase. Moreover, we find that $\langle0{-}n|\mathcal{R}_2|0{-}n\rangle\langle1{+}n|\mathcal{R}_2|1{+}n\rangle=-1$. The localized vacancy states give rise to \mbox{(anti-)}bonding eigenstates at energies $E_0$ and $E_1$. In particular, $|0\rangle$ and $|1\rangle$ have opposite symmetry under $\mathcal{R}_2$. To calculate the exchange coupling as described by Eq.~(\ref{Heff}) it is important that (i) no third state is involved in the superposition of localized states, $\min(\{E_2-E_1,E_0-E_{-1}\})>\Delta$, and that (ii) terms higher than $\mathcal{O}(|t|/U)$ can be neglected, $|t|\ll U$.

\begin{figure}[t!]\centering\includegraphics[width=0.455\textwidth]{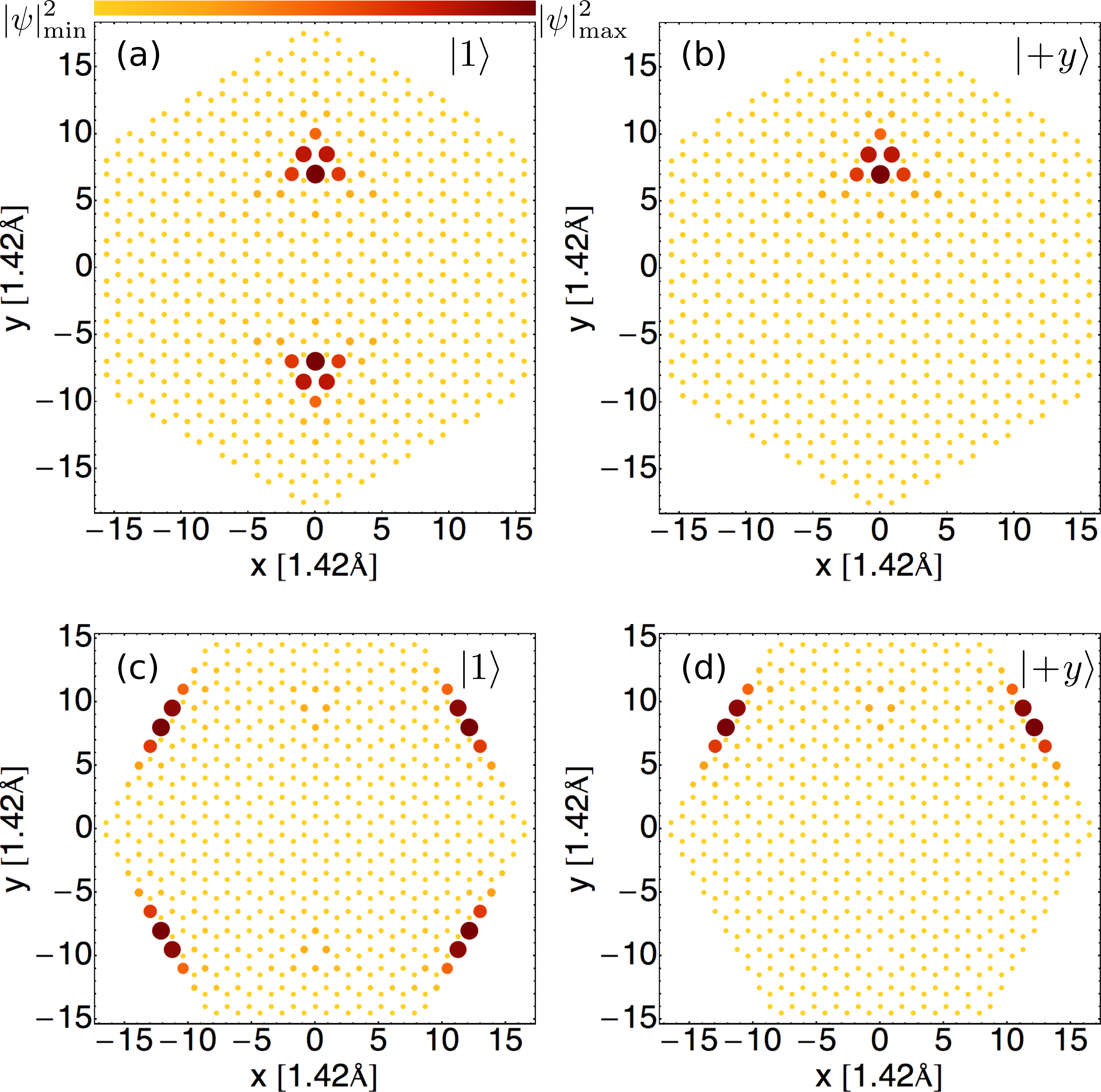}\caption{(Color online) On-site probability densities for (a,b) an ($s{=}5,d{=}8$) armchair terminated flake and (c,d) a ($b{=}10,d{=}10$) zigzag terminated flake. For each, we plot the antibonding energy eigenstates $|1\rangle$ (a,c) as well as the localized vacancy states $|{+}y\rangle$ (b,d). Due to the restriction to nearest neighbor hopping, the on-site probability density of $|0\rangle$ is identical to the one of $|1\rangle$. The probability density of $|{-}y\rangle$ looks similar to the one of $|{+}y\rangle$, yet mirrored about the $x$-axis. There is a significant probability density at the edges for zigzag flakes but not for armchair flakes.}\label{pic5}\end{figure}

\emph{Armchair boundaries.$\,$}
Figure \ref{pic5}(a) and \ref{pic5}(b) show the on-site probability densities of (a) the antibonding eigenstate $|1\rangle$ and (b) the localized state $|\!+\!y\rangle$ for an $(s{=}5,d{=}8)$ nanoflake. The on-site probability density of $|0\rangle$ is the same as that of $|1\rangle$ because the on-site wave functions of these states differ only by a phase. The on-site probability density of the localized state $|\!-\!y\rangle$ is not shown but can be obtained by applying the mirror symmetry $\mathcal{M}_y$ to the probability density of $|{+}y\rangle$. As expected for armchair boundaries, the probability density at the edges is negligible.

The hopping amplitude $|t|$ and the exchange coupling $J$ resulting from the \mbox{(anti-)}bonding states $\{|0\rangle,|1\rangle\}$ are listed in Table~\ref{tac} for $s=1,3,5,7,10,16,22$ and $d=1,2,4,5,7,8,10,11,13,14$ and at vanishing \cite{Bfield} magnetic field. In the top row, we also list the total number of atoms in parentheses. In all other rows, the upper numbers indicate $|t|$ and the lower numbers indicate $J$, both in meV. No results are listed if condition (i) or (ii) is not met and an X is shown if $d$ refers to a lattice site outside the flake. Since $J$ is always positive, it is clear from Eq.~(\ref{Heff}) that the singlet state $|S\rangle$ is favored. The resulting antiferromagnetism should be stable up to the Néel temperature $T_{\rm N}\cong J/k_{\rm B}$, where $k_{\rm B}$ is Boltzmann's constant. The value of $T_{\rm N}$ for a given $(s,d)$ configuration can be obtained by multiplying the according numerical value of $J$ in Table~\ref{tac} with $11.6\,{\rm K}$.
\begin{table}[t!]
\centering
\begin{tabular}{c||r|r|r|r|r|r|r}
\hline\hline
\diagbox{d}{s}&$\begin{matrix}1\\(82)\end{matrix}$&$\begin{matrix}3\\(310)\end{matrix}$&$\begin{matrix}5\\(682)\end{matrix}$&$\begin{matrix}7\\(1198)\end{matrix}$&$\begin{matrix}10\\(2242)\end{matrix}$&$\begin{matrix}16\\(5302)\end{matrix}$&$\begin{matrix}22\\(9658)\end{matrix}$\\
\hline\hline
1& & & & & & & \\
\hline
2& & & & & & & \\
\hline
4&$\begin{matrix}86.997\\\hspace{+0.16cm}\underdash{6.070}\end{matrix}$&$\begin{matrix}104.800\\\hspace{+0.16cm}20.185\end{matrix}$&$\begin{matrix}105.100\\\hspace{+0.16cm}31.746\end{matrix}$& & & & \\
\hline
5&$\begin{matrix}198.193\\\hspace{+0.16cm}48.746\end{matrix}$& & & & & & \\
\hline
7&X&$\begin{matrix}27.165\\\hspace{+0.16cm}\underdash{1.030}\end{matrix}$&$\begin{matrix}38.267\\\hspace{+0.16cm}3.421\end{matrix}$&$\begin{matrix}43.067\\\hspace{+0.16cm}6.041\end{matrix}$&$\begin{matrix}46.041\\\hspace{+0.16cm}9.933\end{matrix}$& & \\
\hline
8&X&$\begin{matrix}109.233\\\hspace{+0.16cm}25.445\end{matrix}$&$\begin{matrix}{\bf107.573}\\\hspace{+0.16cm}{\bf33.687}\end{matrix}$& & & & \\
\hline
10&X&$\begin{matrix}4.003\\\hspace{+0.16cm}\underdash{0.014}\end{matrix}$&$\begin{matrix}14.467\\\hspace{+0.16cm}\underdash{0.406}\end{matrix}$&$\begin{matrix}20.254\\\hspace{+0.16cm}\underdash{1.167}\end{matrix}$&$\begin{matrix}24.890\\\hspace{+0.16cm}2.580\end{matrix}$&$\begin{matrix}28.516\\\hspace{+0.16cm}5.611\end{matrix}$&$\begin{matrix}29.302\\\hspace{+0.16cm}8.469\end{matrix}$\\
\hline
11&X&$\begin{matrix}52.656\\\hspace{+0.16cm}5.595\end{matrix}$&$\begin{matrix}74.052\\16.385\end{matrix}$&$\begin{matrix}73.377\\20.140\end{matrix}$& & & \\
\hline
13&X&X&$\begin{matrix}4.468\\\underdash{0.030}\end{matrix}$&$\begin{matrix}9.568\\\underdash{0.226}\end{matrix}$&$\begin{matrix}14.259\\\hspace{+0.16cm}0.768\end{matrix}$&$\begin{matrix}18.697\\\hspace{+0.16cm}2.222\end{matrix}$&$\begin{matrix}20.439\\\hspace{+0.16cm}3.780\end{matrix}$\\
\hline
14&X&X&$\begin{matrix}47.539\\\hspace{+0.16cm}6.688\end{matrix}$&$\begin{matrix}55.280\\11.682\end{matrix}$&$\begin{matrix}53.900\\14.268\end{matrix}$& & \\
\hline\hline
\end{tabular}
\caption{Results for various armchair terminated flakes with parameters $(s,d)$ (see Fig.~\ref{pic1}). For vanishing magnetic fields \cite{Bfield}, we list the hopping amplitude $|t|$ (upper number) and the exchange coupling $J$ between localized vacancy states (lower number) in ${\rm meV}$. We underline (underdash) $J$ if $J(B){=}0$ can be reached for $B{<}15\text{ T}$ ($B{>}15\text{ T}$). The numbers in boldface correspond to the case shown in Figs.~\ref{pic5}(a), \ref{pic5}(b), and \ref{pic6} (a). We display blank spaces if Eq.~(\ref{Heff}) does not apply (see main text) and “X" if $d$ refers to a lattice site outside the flake.} \label{tac}
\end{table}

\begin{figure}[t!]\centering\includegraphics[width=0.455\textwidth]{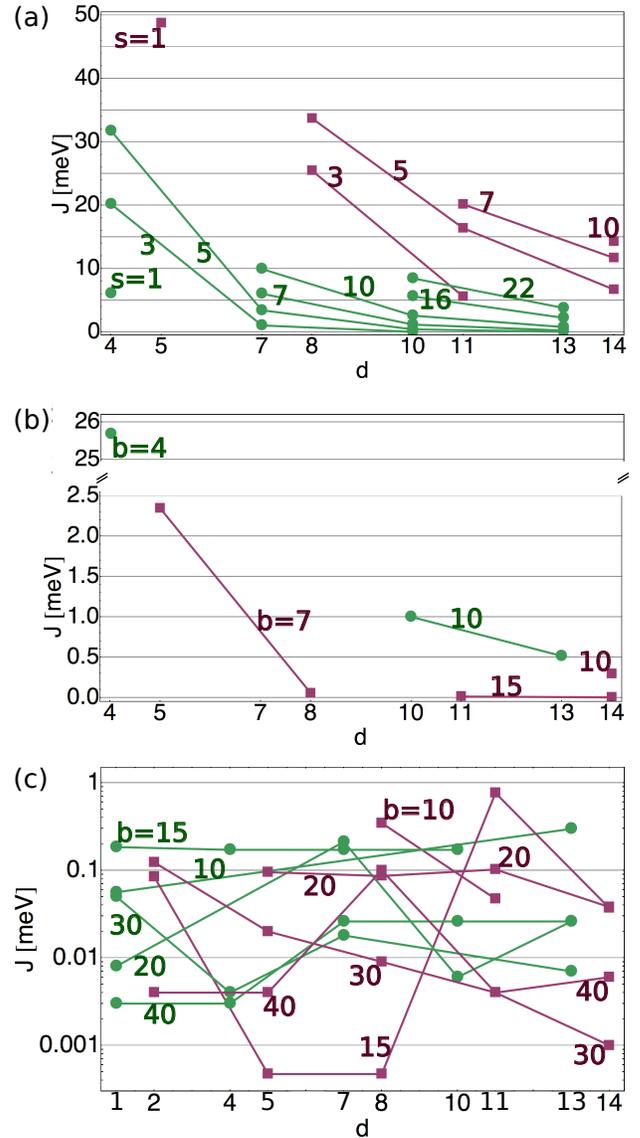}\caption{(Color online) For $t^{(n)}{=}0$ ($n{>}1$), we plot the exchange coupling $J$ for (a) armchair and (b) zigzag terminated flakes. In (c), we plot $J$ for zigzag terminated flakes and $t^{(n)}{\neq}0$ ($n{=}1,2,3$). The values of $J$ are listed in Tables~\ref{tac}, \ref{tzz}, and \ref{tJ}, respectively. Green, circular (magenta, square) markers correspond to $d=1{+}3n$ ($d=2{+}3n$), i.e., vacancy sites $(0,{+}y)$ with a nearest neighbor site at $(0,y{+}a)$ [$(0,y{-}a)$]. Straight lines connect markers that belong to flakes of the same size, e.g., $s{=}3$ (a) or $b{=}15$ [(b) and (c)].}\label{pic4}\end{figure}

For $d=1{+}3n$ ($n{\in}\mathbb{N}$), the lattice site of the vacancy at $(0,+y)$ has a nearest neighbor at $(0,y+a)$ and for $d=2{+}3n$, it has a nearest neighbor at $(0,y-a)$, see Fig.~\ref{pic1}. In Fig.~\ref{pic4} (a), we plot the exchange couplings listed in Table~\ref{tac}. Green, circular (magenta, square) markers correspond to $d=1{+}3n$ ($d=2{+}3n$). Markers that are connected by a straight line belong to flakes of the same size, e.g.,~$s=3$. For armchair boundaries and nearest neighbor hopping only, there is a clear ordering. Both for $d=1{+}3n$ and for $d=2{+}3n$, the exchange coupling decreases with larger vacancy separation $d$ and smaller flake size $s$. The decline of $J$ with increasing $d$ is intuitively clear from Eq.~(\ref{Heff}) as the hopping amplitude $|t|$ decays with increasing separation of the localized states.

\begin{figure}[t!]\centering\includegraphics[width=0.455\textwidth]{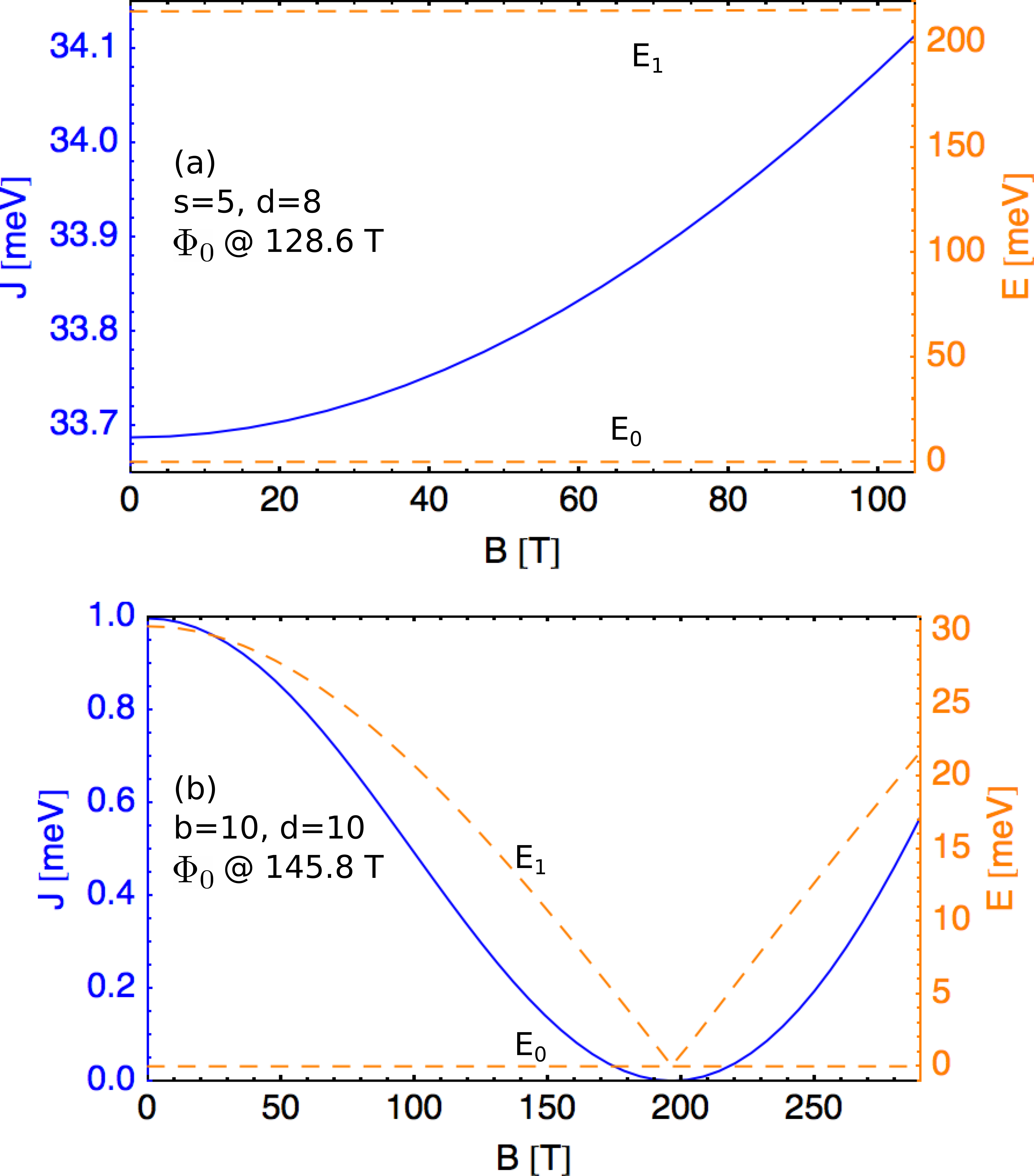}\caption{(Color online) The exchange coupling $J$ (solid blue line and left axis) and the eigenenergies (dashed orange line and right axis) of the corresponding \mbox{(anti-)}bonding states are plotted against a magnetic field perpendicular to the plane of the flake. The flake parameters $(s,d)$ and $(b,d)$, respectively, the energy levels $E_0$ and $E_1$, as well as the magnetic field at which one flux quantum passes through the flake are indicated in the plots. The behavior of $J(B)$ is very specific and depends on flake size ($s$ or $b$), vacancy separation ($d$), edge type, and $|t^{(n)}_{ij}|$ (see also Fig.~\ref{pic3}). (a) and (b) show $J(B)$ for the configurations with numbers in boldface in Tables~\ref{tac} and \ref{tzz}, respectively.}\label{pic6}\end{figure}

The Hamiltonian Eq.~(\ref{hamil}) and hence its spectrum $\{E_n\}$ and the exchange coupling $J$ depend on the magnetic field $B$. In Fig.~\ref{pic6} (a), we plot $J$ and the eigenenergies of the corresponding \mbox{(anti-)}bonding states $\{|0\rangle,|1\rangle\}$ shown in Fig.~\ref{pic5} (a) against $B$. Typically, the properties of an electronic state change for magnetic fields of the order of $B=\Phi_0/A$, where $\Phi_0=h/2e$ is the magnetic flux quantum with Planck's constant $h$ and $A$ is the surface area occupied by the state, which we approximate by the surface area of the flake.

We find that the Coulomb repulsion $U$ depends only weakly on the magnetic field while the splitting $\Delta=E_1-E_0=E_1$ and hence $|t|$ depend strongly on $B$. That is, $J(B)$ is mainly determined by the behavior of $E_1(B)$. Depending on the $(s,d)$ configuration, the exchange coupling $J$ can be tuned over a certain range [Fig.~\ref{pic6} (a)] and if a degeneracy $E_1(B)=E_0=0$ occurs, it is even possible to switch the coupling on ($J>0$) and off ($J=0$) by tuning the system towards or away from the degeneracy. In Tables~\ref{tac}-\ref{tJ}, we underline (underdash) $J$ of those flake configurations, for which a degeneracy, i.e. $J(B)=0$, can be reached with a magnetic field smaller (greater) than 15 T. In Table~\ref{tac}, however, such degeneracies occur only at fields much greater than 15 T.

\emph{Zigzag boundaries.$\,$}
Figures.~\ref{pic5}(c) and \ref{pic5}(d) show on-site probability densities analogous to those in Figs.~\ref{pic5}(a) and \ref{pic5}(b). In contrast to armchair boundaries, the zigzag termination leads to a significant probability density at the flake edges, as expected. Table~\ref{tzz} corresponds to Table~\ref{tac} yet we parametrize the size of zigzag terminated flakes with $b$ instead of $s$ (see Fig.~\ref{pic1}) and here, we use configurations with $b=4,7,10,15,20,30,40$. The flake sizes are chosen such that the number of atoms in the columns of Tables~\ref{tac} and \ref{tzz} roughly match.

The exchange couplings listed in Table~\ref{tzz} are plotted in Fig.~\ref{pic4} (b) in the same way as for armchair boundaries and the available data indicate a decrease of $J$ for larger vacancy separation $d$, as in the armchair case. There are not enough data to draw a conclusion for the behavior of $J$ with respect to the flake size $b$ yet the data points $(b{=}10,d{=}14)$ and $(b{=}15,d{=}14)$ are not in accord with the behavior observed for armchair edges. In Fig.~\ref{pic6} (b), we plot $J$ and the eigenenergies of the corresponding \mbox{(anti-)}bonding states $\{|0\rangle,|1\rangle\}$ shown in Fig.~\ref{pic5} (c) against $B$. As for the armchair edges, degeneracies of \mbox{(anti-)}bonding eigenstates occur only at fields much greater than 15 T.
\begin{table}[t!]
\centering
\begin{tabular}{c||r|r|r|r|r|r|r}
\hline\hline
\diagbox{d}{b}&$\begin{matrix}4\\(94)\end{matrix}$&$\begin{matrix}7\\(292)\end{matrix}$&$\begin{matrix}10\\(598)\end{matrix}$&$\begin{matrix}15\\(1348)\end{matrix}$&$\begin{matrix}20\\(2398)\end{matrix}$&$\begin{matrix}30\\(5398)\end{matrix}$&$\begin{matrix}40\\(9598)\end{matrix}$\\
\hline\hline
1& & & & & & & \\
\hline
2& & & & & & & \\
\hline
4&$\begin{matrix}143.353\\\hspace{+0.16cm}\underdash{25.682}\end{matrix}$& & & & & & \\
\hline
5& &$\begin{matrix}34.886\\\hspace{+0.16cm}\underdash{2.345}\end{matrix}$& & & & & \\
\hline
7&X& & & & & & \\
\hline
8&X&$\begin{matrix}7.044\\\hspace{+0.16cm}\underdash{0.054}\end{matrix}$& & & & & \\
\hline
10&X& &$\begin{matrix}{\bf15.171}\\\hspace{+0.16cm}\underdash{{\bf1.000}}\end{matrix}$& & & & \\
\hline
11&X&X& &$\begin{matrix}1.474\\\underdash{0.012}\end{matrix}$& & & \\
\hline
13&X&X&$\begin{matrix}10.821\\\hspace{+0.16cm}\underdash{0.515}\end{matrix}$& & & & \\
\hline
14&X&X&$\begin{matrix}9.601\\0.293\end{matrix}$&$\begin{matrix}0.724\\\underdash{0.002}\end{matrix}$& & & \\
\hline\hline
\end{tabular}
\caption{Same as Table~\ref{tac}, but for various zigzag terminated flakes with parameters $(b,d)$ (see Fig.~\ref{pic1}) for vanishing magnetic field \cite{Bfield}.} \label{tzz}\end{table}
\subsection{Zigzag terminated flakes with hopping up to 3$^{\text{rd}}$ nearest neighbors}
The restriction to nearest neighbor hopping above leads to a symmetric spectrum with the advantage that the identification of \mbox{(anti-)}bonding eigenstates that lead to localized vacancy states becomes straightforward since these eigenstates lie in the middle of the spectrum at energies $E_0$ and $E_1$. Moreover, this restriction leads to localized vacancy states that reside only in one sublattice \cite{Pereira2006}. In reality, however, all hopping amplitudes $|t^{(n)}_{ij}|$ ($n{>}0$) are finite and as a consequence, the energy spectrum is not symmetric and localized vacancy states reside in both sublattices.

In order to describe a system which is closer to real graphene nanoflakes, we now consider hopping amplitudes up to third nearest neighbors and assume zigzag boundaries, since they are energetically favored in flakes grown by CVD \cite{Luo2011,Phark2011,Subramaniam2012,Hamalainen2011}. The asymmetric energy spectrum makes it challenging to identify the (anti-)bonding eigenstates that are superpositions of the two localized states since these eigenstates lie not necessarily in the middle of the spectrum.

\begin{figure}[t!]\centering\includegraphics[width=0.48\textwidth]{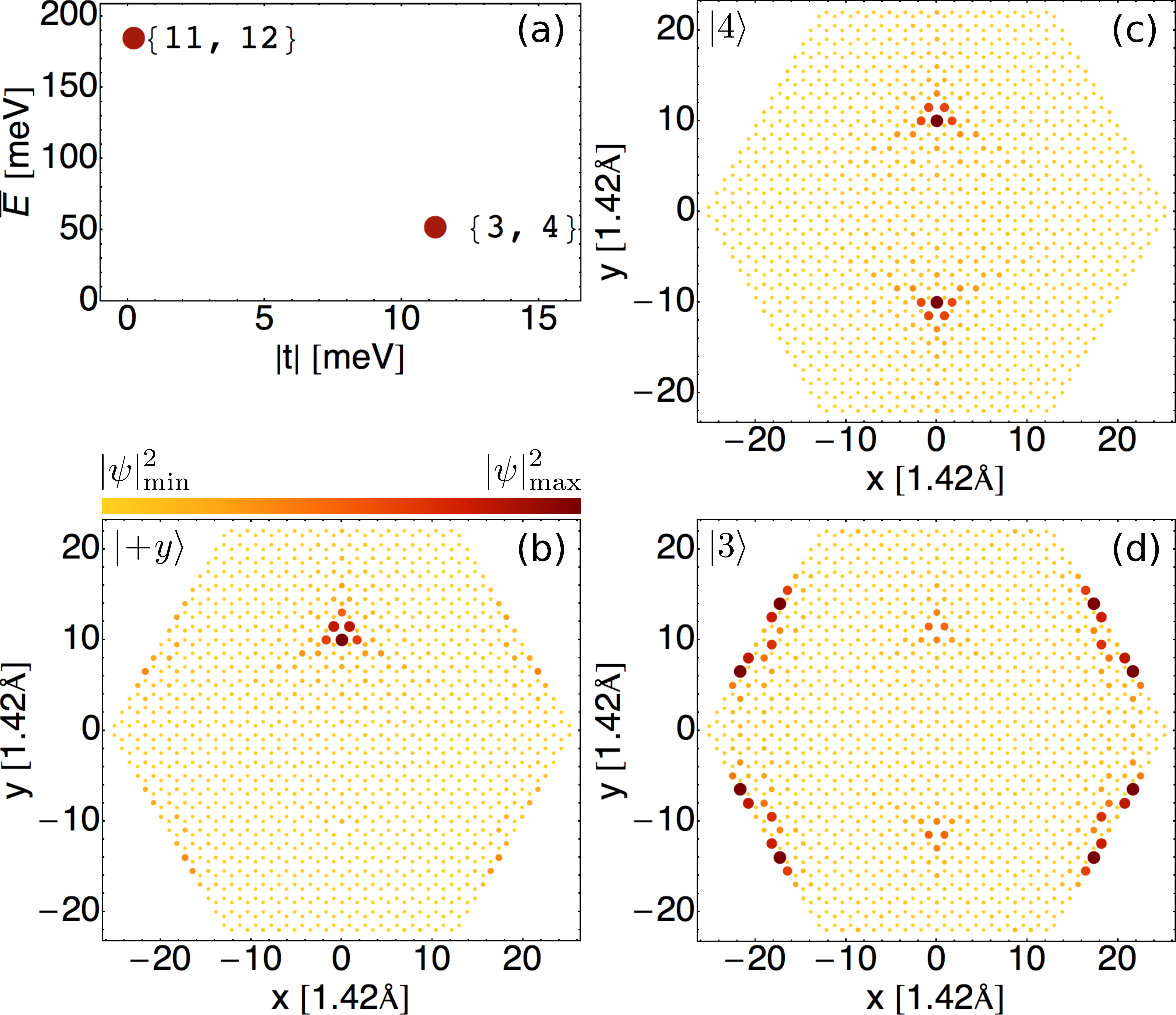}\caption{(Color online) Results for a $(b{=}15,d{=}11)$ flake. (a) Two pairs of \mbox{(anti-)}bonding states satisfy the criteria \mbox{(i)-(v)}. For the corresponding localized states, we plot the energy $\bar{E}$ (with respect to $E_0$) vs.~$|t|$. The hopping amplitude $|t|$ is maximal for the pair $\{|3\rangle,|4\rangle\}$. (b)--(d) On-site probability densities of (b) the localized state $|{+}y\rangle$, (c) of $|4\rangle$, and (d) of $|3\rangle$. The on-site probability density of $|{-}y\rangle$ looks similar to the one of $|{+}y\rangle$, yet mirrored about the $x$ axis.}\label{pic2}\end{figure}
\begin{table}[t!]
\centering
\begin{tabular}{c||r|r|r|r|r}
\hline\hline
\diagbox{$d$}{$b$}&$\begin{matrix}10\\(598)\end{matrix}$&$\begin{matrix}15\\(1348)\end{matrix}$&$\begin{matrix}20\\(2398)\end{matrix}$&$\begin{matrix}30\\(5398)\end{matrix}$&$\begin{matrix}40\\(9598)\end{matrix}$\\
\hline\hline
1&$\,\begin{matrix}3.498\\0.056\end{matrix}$ $(1)\,$&$\begin{matrix}5.150\\\underdash{0.184}\end{matrix}$ $(1)\,$&$\begin{matrix}0.914\\0.008\end{matrix}$ $(1)\,$&$\begin{matrix}1.833\\0.050\end{matrix}$ $(3)\,$&$\begin{matrix}0.426\\\underline{0.003}\end{matrix}$ $(2)$\\
\hline
2&  $ \,$&$\begin{matrix}3.494\\\underdash{0.084}\end{matrix}$ $(3)\,$&  $ \,$&$\begin{matrix}2.937\\\underline{0.123}\end{matrix}$ $(1)\,$&$\begin{matrix}0.457\\\underline{0.004}\end{matrix}$ $(4)\,$\\
\hline
4&  $ \,$&$\begin{matrix}5.048\\\underdash{0.171}\end{matrix}$ $(1)\,$&  $ \,$&$\begin{matrix}0.555\\\underline{0.004}\end{matrix}$ $(1)\,$&$\begin{matrix}0.426\\\underline{0.003}\end{matrix}$ $(3)\,$\\
\hline
5&  $ \,$&$\begin{matrix}0.239\\\hspace{+0.13cm}\underline{0.000}^{\ddag}\end{matrix}\hspace{-0.13cm}$ $(1)\,$&$\begin{matrix}3.216\\0.095\end{matrix}$ $(2)\,$&$\begin{matrix}1.177\\0.020\end{matrix}$ $(3)\,$&$\begin{matrix}0.448\\\underline{0.004}\end{matrix}$ $(5)\,$\\
\hline
7&  $ \,$&$\begin{matrix}5.048\\\underdash{0.171}\end{matrix}$ $(1)\,$&$\begin{matrix}4.922\\\underdash{0.211}\end{matrix}$ $(2)\,$&$\begin{matrix}1.081\\0.018\end{matrix}$ $(6)\,$&$\begin{matrix}1.162\\\underline{0.026}\end{matrix}$ $(2)\,$\\
\hline
8&$\begin{matrix}9.002\\\underdash{0.346}\end{matrix}$ $(1)\,$&$\begin{matrix}0.239\\\hspace{+0.13cm}\underline{0.000}^{\ddag}\end{matrix}\hspace{-0.13cm}$ $(1)\,$&$\begin{matrix}3.076\\\underdash{0.086}\end{matrix}$ $(2)\,$&$\begin{matrix}0.779\\\underline{0.009}\end{matrix}$ $(2)\,$&$\begin{matrix}2.387\\\underline{0.100}\end{matrix}$ $(5)\,$\\
\hline
10&  $ \,$&$\begin{matrix}5.048\\\underdash{0.171}\end{matrix}$ $(2)\,$&$\begin{matrix}0.739\\\underdash{0.006}\end{matrix}$ $(1)\,$&  $ \,$&$\begin{matrix}1.185\\\underline{0.026}\end{matrix}$ $(6)\,$\\
\hline
11&$\begin{matrix}3.301\\\underdash{0.047}\end{matrix}$ $(1)\,$&$\begin{matrix}{\bf11.125}\\\hspace{+0.16cm}{\bf0.762}\end{matrix}$ ${\bf(2)}\,$&$\begin{matrix}3.298\\\underdash{0.102}\end{matrix}$ $(1)\,$&$\begin{matrix}0.526\\\underline{0.004}\end{matrix}$ $(1)\,$&$\begin{matrix}0.460\\\underline{0.004}\end{matrix}$ $(3)\,$\\
\hline
13&$\begin{matrix}7.839\\\underdash{0.297}\end{matrix}$ $(1)\,$&  $ \,$&$\begin{matrix}1.555\\\underdash{0.026}\end{matrix}$ $(3)\,$&$\begin{matrix}0.783\\0.007\end{matrix}$ $(1)\,$&$\begin{matrix}1.162\\\underline{0.026}\end{matrix}$ $(5)\,$\\
\hline
14&  $ \,$&$\begin{matrix}2.428\\0.037\end{matrix}$ $(2)\,$&$\begin{matrix}2.002\\0.038\end{matrix}$ $(3)\,$&$\begin{matrix}0.302\\\underline{0.001}\end{matrix}$ $(2)\,$&$\begin{matrix}0.562\\\underline{0.006}\end{matrix}$ $(5)\,$\\
\hline\hline
\end{tabular}
\caption{Results for various flakes with zigzag edges --- specified by $(b,d)$, see Fig.~\ref{pic1} --- and finite hopping amplitudes up the third nearest neighbors. For vanishing magnetic fields \cite{Bfield}, we list the maximum hopping amplitude $|t|$ (upper number) and the maximal exchange coupling $J$ (lower number) in ${\rm meV}$. We underline (underdash) $J$ if $J(B)=0$ can be reached for $B<15\text{ T}$ ($B>15\text{ T}$). The integer in parentheses behind $|t|$ and $J$ indicates the number of \mbox{(anti-)}bonding pairs that satisfy the criteria \mbox{(i)--(v)}. The numbers in boldface correspond to the case shown in Figs.~\ref{pic2} and \ref{pic3} (a). We display a blank space if Eq.~(\ref{Heff}) does not apply (see main text).\\
$\ddag$ The more accurate value is $0.00047\,{\rm meV}$.} \label{tJ}
\end{table}
For a given flake, we calculate $t$ for all pairs of numerically computed \mbox{(anti-)}bonding states, $\{|n\rangle,|m\rangle\}$, that satisfy three criteria, two of which have been introduced before: (i) the states need to lie next to each other in the spectrum, $m=n+1$ and (ii) they need to have opposite symmetry $\langle n|\mathcal{R}_2|n\rangle\langle m|\mathcal{R}_2|m\rangle=-1$. In addition, (iii) the states  should become accessible via doping in a realistic experiment \cite{Vshift}, $|E_{n,m}-E_0|\leq300\,{\rm meV}$. To calculate the exchange coupling as described by Eq.~(\ref{Heff}), it is important that moreover (iv) no third state is involved in the superposition of localized states, $\min(\{E_n-E_{n-1},E_{m+1}-E_m\})>\Delta$ and that (v) terms higher than $\mathcal{O}(|t|/U)$ can be neglected, $|t|\ll U$.

Figure \ref{pic2} illustrates our results for a $(b{=}15,d{=}11)$ nanoflake. For this flake and at vanishing \cite{Bfield} magnetic field, two pairs of states fulfill the criteria (i)--(v), namely, $\{|3\rangle,|4\rangle\}$ as well as $\{|11\rangle,|12\rangle\}$. In Fig.~\ref{pic2} (a), we plot the hopping amplitude that belongs to the corresponding localized states against the energy $\bar{E}$ of those localized states [Eq.~(\ref{barEt})]. Among the states satisfying the criteria (i)--(v), the states $|3\rangle$ and $|4\rangle$ lead to the highest hopping amplitude, namely, $|t|=11.1\,{\rm meV}$. Figures \ref{pic2}(b)--\ref{pic2}(d) show the on-site probability densities of (b) the localized state $|{+}y\rangle$ and its parent states (c) $|4\rangle$ and (d) $|3\rangle$. The on-site probability density of the localized state $|{-}y\rangle$ is not shown but can be obtained by applying the mirror symmetry $\mathcal{M}_y$ to the on-site probability density of $|{+}y\rangle$.
\begin{figure}[t!]\centering\includegraphics[width=0.48\textwidth]{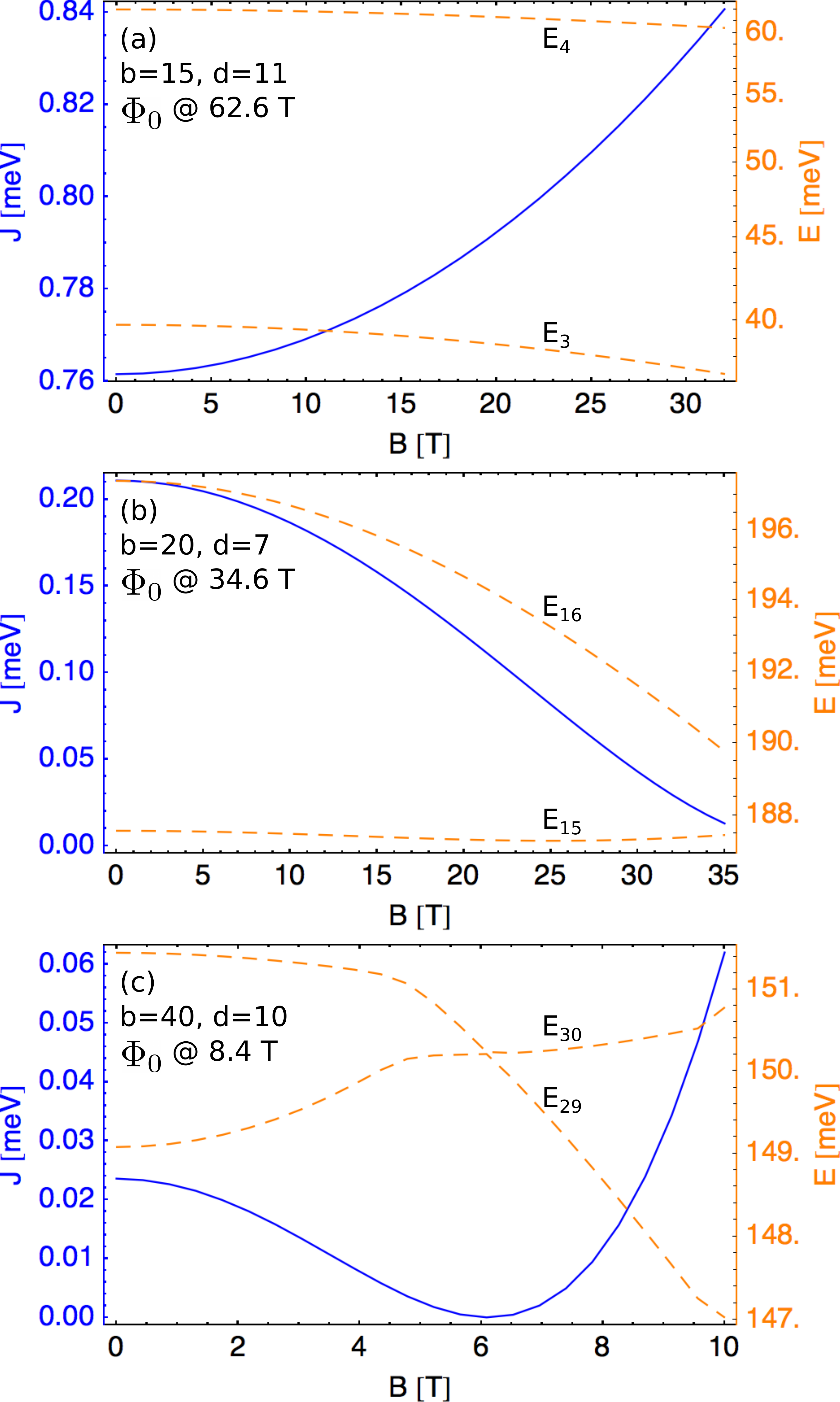}\caption{(Color online) Similar to Fig.~\ref{pic5} (b) but, for the hopping amplitudes between neighbors up to third order. The exchange coupling $J$ (solid blue line) and the energy levels $E_m$ and $E_n$ of the corresponding \mbox{(anti-)}bonding states (dashed orange lines) are shown. Depending on the configuration of the zigzag terminated flake, $J$ can be tuned (a) only weakly or (b) over an order of magnitude, or can be (c) switched off ($J{=}0$) by tuning the spectrum into a degeneracy.}\label{pic3}\end{figure}

For vanishing \cite{Bfield} magnetic fields and any given combination of $b$ and $d$, we now pick the pair of \mbox{(anti-)}bonding states that satisfies the criteria \mbox{(i)--(v)} and which has the highest hopping amplitude. This maximal hopping amplitude $|t|$ and the maximal exchange coupling $J$ are listed in Table~\ref{tJ} for $b{=}10,15,20,30,40$ and $d=1,2,4,5,7,8,10,11,13,14$ in a similar way as in Tables~\ref{tac} and \ref{tzz}: the upper numbers in the table indicate $|t|$ and the lower numbers indicate $J$, both in meV. The ensuing integer in parentheses shows the number of \mbox{(anti-)}bonding pairs that satisfy the criteria (i)--(v) for this combination of $b$ and $d$. The listed values for $J$ are plotted in Fig.~\ref{pic4} (c).

As above, we distinguish between $d{=}1{+}3n$ ($n{\in}\mathbb{N}$) and $d{=}2{+}3n$. The former case leads to a repetitive pattern of $|t|$ and $J$ for e.g., $b{=}15$ and $d{=}4,7,10$ and the latter case applies for e.g., $b{=}15$ and $d{=}5,8$. Such patterns occur for various parameters, yet some of them are concealed in Table~\ref{tJ} because the conditions (i-v) do not apply or the according hopping amplitude $|t|$ is not maximal for a given $(b,d)$ configuration. Throughout a pattern, we find resembling probability densities of the localized states and numerically close but different values for $|t|$ and $J$. In all cases listed in Table~\ref{tJ}, the flake edges play a non-negligible role, see e.g. Fig.~\ref{pic2} (d). This might be the reason why $|t|$ and $J$ vary strongly with respect to $b$ and $d$. For large enough $b$ and $d$, we expect that the influence of $b$ vanishes, and a smooth decay of $|t|$ and $J$ with respect to $d$ occurs; yet we do not reach this regime.

In Fig.~\ref{pic3}, we plot $J$ and the eigenenergies of the corresponding \mbox{(anti-)}bonding states $\{|E_n\rangle,|E_m\rangle\}$ against $B$ for three different flake configurations. As before, $J$ is mainly determined by the energy splitting of the \mbox{(anti-)}bonding eigenstates, $\Delta=E_m-E_n$. Depending on the $(b,d)$ configuration, the exchange coupling $J(B)$ can be tuned over a certain range [Figs.~\ref{pic3}(a) and \ref{pic3}(b)]; and if a degeneracy $E_m(B)=E_n(B)$ occurs, it is even possible to switch the coupling on ($J>0$) and off ($J=0$) by tuning the system towards or away from the degeneracy [Fig.~\ref{pic3}(c)]. In Table~\ref{tJ}, we underline (underdash) $J$ of those flake configurations, for which a degeneracy, i.e., $J(B)=0$, can be reached with a magnetic field smaller (greater) than 15 T.
\section{Conclusion and Outlook}
We have set up a tight-binding model for hexagonal graphene nanoflakes with zigzag edges and two vacancies at positions $\vec{r}_{\rm vac}=(0,\pm y)$. Symmetry allows us to infer the explicit form of the localized vacancy states from the bonding and antibonding eigenstates. This system is a realization of a DQD. In the weak tunneling regime, the triplet $|T_0\rangle$ and the singlet $|S\rangle$ are split by the exchange coupling $J$ and their dynamics decouples from other spin states. A perpendicular magnetic field is included in the tight-binding model via a Peierls phase and can be used to tune $J$ by orders of magnitude, depending on the flake configuration.

We consider flakes with armchair and zigzag edges where we restrict the tight-binding model to hopping between nearest neighbors. Motivated by experiments on CVD grown graphene nanoflakes, we also discuss zigzag terminated flakes where hopping amplitudes up to third nearest neighbors are taken into account. In the former two cases, the calculation of $J$ is straightforward. In the latter case, we have calculated $J$ for states that can be reached via doping leading to a shift of the chemical potential by less than $\pm300\,{\rm meV}$ and that satisfy further criteria described above. For flakes with armchair edges, the exchange coupling decays with increasing separation of the vacancies. It remains unclear whether such behavior also applies for zigzag terminated flakes, where edge states play a significant role.

Due to the dependence of $J(B)$ on the perpendicular magnetic field it is possible to tune the system into a degeneracy where $J{=}0$. This in-situ tunability of the exchange coupling can be very useful for spintronics and quantum-information-related applications because it allows the modification of the magnetic properties without preparing  a new device. The ground-state spin configuration is antiferromagnetic. Depending on the lattice configuration, we have found Néel temperatures from below $4\,$K to beyond room temperature, which allows experimental testing of our results. Ferromagnetic ordering, $J{<}0$, is conceivable by including non-local Coulomb interaction \cite{Burkard1999}. Our calculation can be extended to include spin-orbit coupling or additional potentials that model a Moiré pattern or boundary effects. Assigning both vacancies to the same sublattice results in reduced symmetry. These cases might be treatable with a modified, less- symmetry-dependent calculation.
\section{Acknowledgements}
We thank the European Science Foundation and the Deutsche Forschungsgemeinschaft (DFG) for support within the EuroGRAPHENE project CONGRAN and the DFG for funding within SFB 767 and FOR 912.


\begin{thebibliography}{}
\bibitem{Bfield}
For $B=0$, the Hamiltonian does not only possess the symmetry $\mathcal{R}_2$ but also $\mathcal{M}_x$ and $\mathcal{M}_y$, as described below Eq.~(\ref{phase}). To avoid degeneracies, we use a vanishingly small magnetic field of $10^{-7}\,{\rm T}$ instead of $B=0$.
\end{thebibliography}
\end{document}